\documentclass[12pt]{article}
\usepackage{graphicx}

\def\mystrut{\vrule height 3.5ex depth 2.2ex width 0pt}
\def\downstrut{\vrule height 1ex depth 3.0ex width 0pt}
\def\medstrut{\vrule height 1ex depth 1.0ex width 0pt}
\def \beq{\begin{equation}}
\def \eeq{\end{equation}}
\def\bea{\begin{eqnarray}}
\def\eea{\end{eqnarray}}
\def\eqref#1{(\ref{#1})}
\def\red{
}
\def\black{
}

\def\URLtilde{\lower0.2em\hbox{$\tilde{\phantom{a}}$}}
\def\mycomm#1{\hfill\break\strut\kern-3em{\red\tt ====> #1\black}\hfill\break}
\def\mycommNL#1{\strut\kern-4em{\red\tt ====> #1\black}\hfill\break}

\def\gray{\special{ps: 0.40 setgray}}
\def\black{\special{ps: 0.0 setgray}}

\newcommand{\mydraft}{
\newcount\timecount
\newcount\hours \newcount\minutes  \newcount\temp \newcount\pmhours

\hours = \time
\divide\hours by 60
\temp = \hours
\multiply\temp by 60
\minutes = \time
\advance\minutes by -\temp
\def\hour{\the\hours}
\def\minute{\ifnum\minutes<10 0\the\minutes
    \else\the\minutes\fi}
\def\clock{
\ifnum\hours=0 12:\minute\ AM
\else\ifnum\hours<12 \hour:\minute\ AM
\else\ifnum\hours=12 12:\minute\ PM
    \else\ifnum\hours>12
     \pmhours=\hours
     \advance\pmhours by -12
     \the\pmhours:\minute\ PM
     \fi
    \fi
\fi
\fi
}
\def\fullclock{\hour:\minute}
\begin{centering}
\gray
\font\Hugett  =cmtt12 scaled\magstep4
\hbox{\Hugett Draft:\today,\clock}
\black
\end{centering}
\vskip -1.7cm
$\phantom{a}$
} 

\begin{document}
\thispagestyle{empty}
\rightline{EFI 07-23}
\rightline{TAUP 2863/07}
\rightline{WIS/14/07-AUG-DPP}
\rightline{ANL-HEP-PR-07-58}
\vskip3cm

\centerline{\large \bf Predictions for masses of bottom baryons}
\bigskip

\centerline{Marek Karliner$^a$, Boaz Keren-Zur$^a$, Harry J. Lipkin$^{a,b,c}$,
and Jonathan L. Rosner$^d$}
\medskip

\centerline{$^a$ {\it School of Physics and Astronomy}}
\centerline{\it Raymond and Beverly Sackler Faculty of Exact Sciences}
\centerline{\it Tel Aviv University, Tel Aviv 69978, Israel}
\medskip

\centerline{$^b$ {\it Department of Particle Physics}}
\centerline{\it Weizmann Institute of Science, Rehovoth 76100, Israel}
\medskip

\centerline{$^c$ {\it High Energy Physics Division, Argonne National
Laboratory}}
\centerline{\it Argonne, IL 60439-4815, USA}
\medskip

\centerline{$^d$ {\it Enrico Fermi Institute and Department of Physics}}
\centerline{\it University of Chicago, 5640 S. Ellis Avenue, Chicago, IL
60637, USA}
\bigskip
\strut
\bigskip
\strut
\bigskip

\centerline{\bf ABSTRACT}
\bigskip

\begin{quote}
The recent observation of $\Sigma_b^{\pm}$ ($uub$ and $ddb$) and $\Xi_b^-$
$(dsb)$ baryons at the Tevatron within 2~MeV of our theoretical predictions
provides a strong motivation for applying the same theoretical approach,
based on modeling the color hyperfine interaction, to predict the masses of
other bottom baryons which might be observed in the foreseeable future.
For S-wave $qqb$ states we predict 
$M(\Omega_b) = 6052.1 \pm 5.6$ MeV,
$M(\Omega^*_b) = 6082.8 \pm 5.6$ MeV, 
and $M(\Xi_b^0) = 5786.7 \pm 3.0$ MeV.
For states with one unit of orbital angular momentum between the $b$ quark
and the two light quarks we predict $M(\Lambda_{b[1/2]}) = 5929 \pm 2$ MeV,
$M(\Lambda_{b[3/2]}) = 5940 \pm 2$ MeV, $M(\Xi_{b[1/2]}) = 6106 \pm 4$ MeV,
and $M(\Xi_{b[3/2]}) = 6115 \pm 4$ MeV.
\end{quote}
\vfill
\leftline{PACS codes: 14.20.Mr, 12.40.Yx, 12.39.Jh, 11.30.Hw}

\vfill\eject

\begin{section}{Introduction}

There has been noteworthy experimental progress recently in the identification
of baryons containing a single $b$ quark.  The CDF Collaboration has seen the
states $\Sigma_b^\pm$ and $\Sigma_b^{*\pm}$ \cite{CDFsigmab}, while both D0
\cite{Abazov:2007ub} and CDF \cite{CDFxib} have observed the $\Xi_b^-$.

The constituent quark model has been remarkably successful in predicting the
masses of these states \cite{earlier}-\cite{Karliner:2006ny}.  Most recently
the careful accounting for wave function effects in the hyperfine interaction
\cite{Karliner:2007xib} has permitted the prediction of the $\Xi^-_b$ mass
within a few MeV of observation.

All predictions need an input for the mass difference $m_b-m_c$
between the $b$ and $c$ quarks. That the value of $m_b-m_c$ obtained from
hadrons containing $b$ and $c$ quarks depends upon the flavors of the
spectator quarks was noted in Ref.~\cite{Karliner:2003sy} where Table I
shows that the value is the same for mesons and baryons not containing
strange quarks but different when obtained from $B_s$ and $D_s$ mesons.
Some reasons for this difference were noted and the issue still requires
further investigation.

The new CDF mass measurement~\cite{CDFxib} of the baryon $\Xi^-_b$ confirms
the prediction \cite{Karliner:2007xib} which uses the value of $m_b-m_c$
obtained from $B_s$ and $D_s$ meson masses. 
Therefore in our present analysis we use this value of
of $m_b-m_c$, as well as a very close value obtained from the $\Xi_b - \Xi_c$
mass difference.
 Since these values are  about 10 MeV lower than the
value obtained \cite{Karliner:2003sy} from nonstrange hadrons, our 
predictions 
are lower than other predictions~\cite{Jenkins:1996de,Ebert:2005xj} which
use nonstrange hadron masses as inputs.

In this model the mass of a hadron is given by the sum of the constituent quark
masses plus the color-hyperfine (HF) interactions:
\begin{equation}
V^{HF}_{ij}=v\frac{\vec{\sigma_i}\cdot\vec{\sigma_j}}{m_im_j}
\langle\delta(r_{ij})\rangle
\end{equation}
where the $m_i$ is the mass of the $i$-th constituent quark, $\sigma_i$ its
spin, $r_{ij}$ the distance between the quarks and $v$ is the interaction
strength. We shall neglect the mass differences between $u$ and $d$
constituent quarks, writing $u$ to stand for either $u$ or $d$. All the hadron
masses (except the ones given in Sec.\ \ref{sec_xib_isospin}) are for
isospin-averaged baryons.

Two interesting observations, based on a study of the hadronic spectrum, lead
to improved predictions for the $b$ baryons. The first is that the effective
mass of the constituent quark depends on the spectator quarks
\cite{Karliner:2003sy,Karliner:2007xib}, and the second is an effective
supersymmetry \cite{Karliner:2006ny} -- a resemblance between mesons and
baryons where the anti-quark is replaced by a diquark \cite{Lichtenberg:1989ix}.

In this paper we extend the same methodology to obtain predictions for the
masses of additional baryonic states containing the $b$ quark that will be
experimentally accessible in the foreseeable future.
\end{section}

\newpage

\def\mystrut{\vrule height 3.5ex depth 2.2ex width 0pt}
\section{$\Omega_b$ mass prediction}
\strut\vskip-2em 
\begin{table}[h]
\caption{Hadron masses used in the calculation of the $\Omega_b$ mass prediction
\label{tab:masses}}
\begin{center}
\begin{tabular}{l c} \hline \hline
Splitting                             & Value (MeV) \\ \hline
$M(\Omega_c)$                         & $2697.5\phantom{0}\pm 2.6$ \\
$M(\Omega_c^*)$                       & $2768.3\phantom{0}\pm 3.0$ \\
$M(\Omega_c^*)-M(\Omega_c)$           & $\phantom{00}70.8\phantom{0}\pm 1.5$ \\
$M(D_s)$                              & $1968.49 \pm 0.34$ \\
$M(D_s^*)$                            & $2112.3\phantom{0}\pm 0.5$ \\
$M(B_s)$                              & $5366.1\phantom{0}\pm 0.6$ \\
$M(B_s^*)$                            & $5412.0\phantom{0}\pm 1.2$ \\
$M(B_s^*)-M(B_s)$                     & $\phantom{00}45.9\phantom{0} \pm 1.2$ \\
$M(\Xi_c^0)$                          & $2471.0\phantom{0} \pm 0.4$ \\ 
$M(\Xi_b^-)$                          & $5792.9\phantom{0} \pm 3.0$ \\
\hline
\hline
\end{tabular}
\end{center}
\end{table}
Taking the approach implemented in \cite{Karliner:2007xib} for the prediction
of the $\Xi_b$ mass, the spin averaged mass of $\Omega_b$ can be obtained by
extrapolation from available data for $\Omega_c$ and a correction based on
strange meson masses, as listed in Table \ref{tab:masses}:
\begin{eqnarray}
\label{Omegab-spin-ave}
M(\widetilde{\Omega_b})&\equiv&
\frac{2M(\Omega_b^*)+M(\Omega_b)}{3}=\frac{2M(\Omega_c^*)+M(\Omega_c)}{3}
+{(m_b-m_c)\medstrut}_{B_s-D_s}
\downstrut 
\\
 \nonumber
&=&\frac{2M(\Omega_c^*)+M(\Omega_c)}{3}
+\frac{3M(B_s^*)+M(B_s)}{4}-\frac{3M(D_s^*)+M(D_s)}{4} 
\downstrut 
\\
\nonumber
&=&6068.9\pm 2.4~\textnormal{MeV}
\end{eqnarray}
where $M(\widetilde{X})$ denotes the spin-averaged mass
 that cancels out the hyperfine interaction between the
heavy quark and the diquark containing lighter quarks.
\downstrut

The HF splitting can be estimated as follows:
\begin{eqnarray}
M(\Omega_b^*)-M(\Omega_b) & = &(M(\Omega_c^*)-M(\Omega_c))\frac{m_c}{m_b} =
24.0 \pm 0.7~\textnormal{MeV}~,
\end{eqnarray}
where we have used the experimental mass difference
\cite{Aubert:2006je} $M(\Omega_c^*) - M(\Omega_c) =\break
70.8 \pm 1.0 \pm 1.1~{\rm MeV} = 70.8 \pm 1.5~{\rm MeV}$
with $m_b/m_c$ taken to be $2.95 \pm 0.06 $, as discussed in the Appendix.
This gives the following mass predictions:
\begin{equation}
\Omega_b^* = 6076.9\pm 2.4~\rm{MeV}; ~ ~ ~ \Omega_b = 6052.9\pm2.4~\rm{MeV}
\end{equation}

Taking into account the wavefunction correction as described in
\cite{KerenZur:2007vp}, one must add the following correction to
the spin averaged mass:
\begin{eqnarray}
v\Bigg[\frac{\langle \delta(r_{ss}) \rangle_{\Omega_b}}{m_s^2}
-\frac{\langle \delta(r_{ss}) \rangle_{\Omega_c}}{m_s^2}\Bigg]
&=& v\frac{\langle \delta(r_{ss}) \rangle_{\Omega_c}}{m_s^2}
\Bigg[\frac{\langle \delta(r_{ss}) \rangle_{\Omega_b}}{\langle \delta(r_{ss})
\rangle_{\Omega_c}}-1\Bigg] \nonumber \\
&\approx&(50\pm10)\Bigg[\frac{\langle \delta(r_{ss}) \rangle_{\Omega_b}}
{\langle \delta(r_{ss}) \rangle_{\Omega_c}}-1\Bigg]=2.0\pm1.1~\rm{MeV}
\label{WF-corr}
\end{eqnarray}
where the contact probability ratio was computed using variational methods
\beq
\frac{\langle \delta(r_{ss}) \rangle_{\Omega_b}}{\langle \delta(r_{ss}) \rangle_{\Omega_c}}=1.04\pm0.02~,
\eeq
and we used the following calculation to evaluate the strength of the $ss$ HF
interaction:
\begin{eqnarray}
50~\rm{MeV}&\approx&M(\Omega)+\frac{1}{4}(2M(\Xi_c^*)+M(\Xi_c')+M(\Xi_c))
\nonumber
\\
&&-\frac{1}{3}(2M(\Xi^*)+M(\Xi))-\frac{1}{3}(2M(\Omega_c^*)+M(\Omega_c))
=\nonumber \\
&=&\Bigg(3m_s+3v\frac{\langle \delta(r_{ss})
\rangle_{\Omega}}{m_s^2}\Bigg)+\Bigg(m_u+m_s+m_c\Bigg) \nonumber \\
&&-\Bigg(2m_s+m_u+v\frac{\langle \delta(r_{ss}) \rangle_{\Xi}}{m_s^2}\Bigg)
-\Bigg(2m_s+m_c+v\frac{\langle \delta(r_{ss}) \rangle_{\Omega_c}}{m_s^2}\Bigg)
\nonumber\\
&\approx& v\frac{\langle \delta(r_{ss}) \rangle}{m_s^2}
\end{eqnarray}

\subsubsection*{An alternate derivation of the $\Omega_b$ mass from the
$\Xi_b - \Xi_c$ mass difference}

Thanks to new measurements of the $\Xi_b^-$ mass \cite{Abazov:2007ub,CDFxib}, 
we now have another way of estimating the spin-averaged $\Omega_b$ mass.
Following the approach in Ref.~\cite{Karliner:2007xib}
the $\Xi_b^- - \Xi_c^0$ mass difference can be schematically written as 
\beq
\begin{array}{ccccccc}
M(\Xi_b^-)-M(\Xi_c^0)&=& (m_b - m_c) &+& \hbox{(wavefunction correction)}
&+& \hbox{(EM correction)}
\\
\\
&=& (m_b - m_c) &+& ({-}4\pm4)~{\rm MeV}
&+&(V^{EM}_{bsd}-V^{EM}_{csd})
\\
\\
\end{array}
\eeq
where the value of the wave function correction is taken from
\cite{Karliner:2007xib} and the last term denotes the EM interactions of
the relevant quarks.\downstrut

Similarly, the spin-averaged $\Omega_b-\Omega_c$ mass difference can be
written as
\beq
\begin{array}{ccccccc}
M(\widetilde{\Omega_b})-M(\widetilde{\Omega_c})
&=& (m_b - m_c) &+& \hbox{(wavefunction correction)}
&+& \hbox{(EM correction)}
\\
\\
&=& (m_b - m_c) &+& (2.0\pm1.1)~{\rm MeV}
&+&(V^{EM}_{bss}-V^{EM}_{css})
\\
\\
\end{array}
\eeq
where the wave-function correction is given in Eq.~\eqref{WF-corr}.
\downstrut

Since the $b$ and $s$ quarks have the same charge, the EM contribution
\hbox{$V^{EM}_{bss}-V^{EM}_{css}$} to the $\Omega_b - \Omega_c$ mass difference
is almost the same as the EM contribution \hbox{$V^{EM}_{bsd}-V^{EM}_{csd}$}
to the $\Xi_b^- - \Xi_c^0$ mass difference, modulo a negligible correction from
the change in the mean radius of the relevant baryons. We then immediately
obtain
\beq
M(\widetilde{\Omega_b})-M(\widetilde{\Omega_c}) =
M(\Xi_b^-)-M(\Xi_c^0)+(6.0\pm4.1)~{\rm MeV}
\eeq
which leads to
\beq
M(\widetilde{\Omega_b})=6072.6\pm 5.6~\textnormal{MeV}
\label{Omegab-from-Xi}
\eeq
to be compared with 
$M(\widetilde{\Omega_b})=6070.9\pm 2.7~\textnormal{MeV}$ 
from Eqs.~\eqref{Omegab-spin-ave} and \eqref{WF-corr}.
\downstrut

The consistency of these two estimates, based on different
experimental inputs, is a strong indication that 
both the central values and the error estimates are reliable.
Moreover, the estimate in Eq.~\eqref{Omegab-from-Xi} includes EM corrections,
while the estimate Eqs.~\eqref{Omegab-spin-ave} does not, thus indicating
that the EM corrections are likely to be smaller than our error estimate.
Consequently, in the following we use the estimate
\eqref{Omegab-from-Xi}.

\subsubsection*{Wave function correction to the hyperfine splitting}

We must also compute the correction to the HF splitting
\begin{eqnarray}
M(\Omega_b^*)-M(\Omega_b)&=&
(M(\Omega_c^*)-M(\Omega_c))\frac{m_c}{m_b}\frac{\langle \delta(r_{bs})
\rangle_{\Omega_b}}{\langle \delta(r_{cs}) \rangle_{\Omega_c}}
=30.7 \pm1.3~\textnormal{MeV}
\end{eqnarray}
where we used
\beq
\frac{\langle \delta(r_{bs}) \rangle_{\Omega_b}}{\langle \delta(r_{cs})
 \rangle_{\Omega_c}}=1.28\pm0.04~,
\eeq
leading to the following predictions:
\begin{equation}
\Omega_b^* = 6082.8\pm5.6~\rm{MeV}; ~ ~ ~ \Omega_b = 6052.1\pm5.6~\rm{MeV}
\label{omegab_pred}
\end{equation}

\subsubsection*{An alternative derivation of HF splitting from effective supersymmetry}
An alternative approach to estimate the HF splitting is to use the effective
meson-baryon supersymmetry discussed in \cite{Karliner:2006ny} and
apply it to the case of hadrons related by changing a strange antiquark
$\bar s$ to a doubly strange $ss$ diquark coupled to spin $S = 1$:

\begin{equation}
\begin{array}{ccccccc}
\displaystyle
{{M(\Omega_b^*) - M(\Omega_b)}\over{M(B_s^*)-M(B_s)}} &=&
\displaystyle
{{M(\Omega_c^*){-}M(\Omega_c)}\over{M(D_s^*){-}M(D_s)}}
&=&
\displaystyle
{{M(\Xi^*){-}M(\Xi)}\over{M(K^*){-}M(K)}}
\strut\kern-5em\strut
\label{eq:newpredo}
\\
\\
[4pt]
&\approx &
0.49 \pm 0.01
&\approx&
0.54
\end{array}
\end{equation}

\begin{eqnarray}
\Omega_b^*-\Omega_b=(B_s^*-B_s)(0.52\pm0.02)=23.9\pm1.1~\textnormal{MeV}
\end{eqnarray}
This gives \begin{equation} 
\Omega_b^* = 6076.8\pm2.4~\rm{MeV};
~ ~ ~ 
\Omega_b = 6053.0\pm2.5~\rm{MeV}~. 
\end{equation}

The main difference between these predictions and the ones given in the
past~\cite{Jenkins:1996de,Ebert:2005xj} 
is the use of masses of hadrons containing strange quarks
~\cite{Karliner:2007xib}, rather than $\Lambda_b$ and $\Lambda_c$ masses,
to obtain the quark mass difference $m_b-m_c$. We also take into account wave
function corrections which influence the hyperfine splitting between
$\Omega_b^*$ and $\Omega_b$.  The net
result is that we predict substantially lower masses for $\Omega_b$
than both Ref.~\cite{Jenkins:1996de}: $M(\Omega_ b)=6068.7\pm11.1$ MeV, 
and Ref.~\cite{Ebert:2005xj}: $M(\Omega_ b)=6065$ MeV.
Our predicted hyperfine splitting $M(\Omega_b^*) - M(\Omega_b) = 30.7 \pm 1.3$
MeV (when wave function effects are included) is also larger  than those of
Refs.\ \cite{Jenkins:1996de} (14.5 MeV) and \cite{Ebert:2005xj} (23 MeV).

\section{$\Xi_b$ isospin splitting}
\label{sec_xib_isospin}
The $\Xi_b^0$ mass is expected to be measured by the CDF collaboration through
the channel $\Xi_b^0 \to \Xi_c^+ \pi^-$, where $\Xi_c^+ \to \Xi^- \pi^+ \pi^+$,
$\Xi^- \to \Lambda \pi^-$, and $\Lambda \to p \pi^-$ \cite{Litvintsev}.

The source for the isospin splitting ($\Delta I $) is the difference in the
mass and charge of the $u$ and $d$ quarks. These differences affect the hadron
mass in four ways \cite{Rosner:1998zc}: they change the constituent quark
masses ($\Delta M = m_d - m_u$), the Coulomb interaction ($V^{EM}$), and the
spin-dependent interactions -- both magnetic and chromo-magnetic ($V^{spin}$).
One can obtain a prediction for the $\Xi_b$ isospin splitting by extrapolation
from the $\Xi$ data, which has similar structure as far as EM interactions are
concerned (note that for $\Xi_b$ there are no spin-dependent interactions
between the heavy quark and the $su$ diquark which is coupled to spin zero):
\begin{eqnarray}
\Delta I(\Xi^*)&=&\Delta M + \Bigg[V^{EM}_{ssd} - V^{EM}_{ssu}\Bigg]
 +2 \Bigg[V^{spin}_{ds} - V^{spin}_{us}\Bigg]= 3.20\pm0.68~\textnormal{MeV} \\
\Delta I(\Xi)&=&\Delta M + \Bigg[V^{EM}_{ssd} - V^{EM}_{ssu}\Bigg]
 -4 \Bigg[V^{spin}_{ds} - V^{spin}_{us}\Bigg]= 6.85\pm0.21~\textnormal{MeV}\\
\strut\kern-5em
\Rightarrow\Delta I(\Xi_b)&=&\Delta M + \Bigg[V^{EM}_{ssd} - V^{EM}_{ssu}\Bigg]
 -3 \Bigg[V^{spin}_{ds} - V^{spin}_{us}\Bigg]\\ \nonumber
&=&\frac{2\Delta I(\Xi^*)+\Delta I(\Xi)}{3}
 +\frac{\Delta I(\Xi)-\Delta I(\Xi^*)}{2} \\ \nonumber
&=&\frac{\Delta I(\Xi^*)+5\Delta I(\Xi)}{6} \\ \nonumber
&=&6.24\pm0.21~\textnormal{MeV}
\end{eqnarray}
With the observed value \cite{CDFxib} $M(\Xi_b^-) = (5792.9 \pm 2.5 \pm 1.7)$
MeV (the error from the D0 experiment is considerably
larger \cite{Abazov:2007ub}) and this estimate, we predict
$M(\Xi_b^0) = 5786.7 \pm 3.0$ MeV.

Another option is to use $\Xi_c$, which has the same spin-dependent
interactions, as a starting point:
\begin{eqnarray}
\Delta I(\Xi_c)&=&\Delta M + \Bigg[V^{EM}_{csd} - V^{EM}_{csu}\Bigg] -3 \Bigg[V^{spin}_{ds} - V^{spin}_{us}\Bigg]= 3.1\pm0.5~\textnormal{MeV} \\
\Rightarrow\Delta I(\Xi_b)&=&\Delta M + \Bigg[V^{EM}_{ssd} - V^{EM}_{ssu}\Bigg] -3 \Bigg[V^{spin}_{ds} - V^{spin}_{us}\Bigg]\\ \nonumber
&=&\Delta I(\Xi_c)+\Bigg[V^{EM}_{ssd} - V^{EM}_{ssu}\Bigg]-\Bigg[V^{EM}_{csd} - V^{EM}_{csu}\Bigg]\\ \nonumber
&=&\Delta I(\Xi_c)+\frac{2\Delta I(\Xi^*)+\Delta I(\Xi)}{3}-\frac{2\Delta I(\Xi_c^*)+\Delta I(\Xi_c')+\Delta I(\Xi_c)}{4}\\ \nonumber
&=&6.4\pm1.6~\textnormal{MeV}
\end{eqnarray}
We summarize the isospin splittings which have been used in these calculations
in Table \ref{tab:split}.  All masses have been taken from the 2007 updated
tables of the Particle Data Group \cite{PDG07}, and all values of $\Delta I$
are defined as $M$(baryon with $d$ quark) - $M$(baryon with $u$ quark).


\begin{table}
\caption{Isospin splittings $\Delta I$ used in calculating
$\Delta I(\Xi_b) \equiv M(\Xi_b^-) - M(\Xi_b^0)$.
\label{tab:split}}
\begin{center}
\begin{tabular}{l l} \hline \hline
Splitting & Value (MeV) \\ \hline
$\Delta I(\Xi)$     & $\phantom{-}6.85 \pm 0.21$ \\
$\Delta I(\Xi^*)$   & $\phantom{-}3.20 \pm 0.68$ \\
$\Delta I(\Xi_c)$   &  $\phantom{-}3.1\phantom{0} \pm 0.5$  \\
$\Delta I(\Xi'_c)$  &  $\phantom{-}2.3\phantom{0} \pm 4.24$ \\
$\Delta I(\Xi^*_c)$ & ${-}0.5\phantom{0} \pm 1.84$ \\ \hline \hline
\end{tabular}
\end{center}
\end{table}

\section{$\Lambda_b$ and $\Xi_b$ orbital excitations}
\strut\vskip-2em 
\begin{table}[h]
\caption{Masses of $\Lambda$ and $\Xi$ baryon ground states and orbital
excitations~\cite{PDG07}.
\label{tab:orbital}}
\begin{center}
\begin{tabular}{l c c c c} \hline \hline
         & $\Lambda$              &$\Lambda_c$         &$\Xi_c^+$           &
$\Xi_c^0$           \\ \hline
$M(1/2^+)$ & $1115.683 \pm 0.006$  & $2286.46 \pm 0.14$ &$2467.9  \pm 0.4$   &
$2471.0   \pm 0.4$  \\
$M(1/2^-)$ & $1406.5   \pm 4.0$    & $2595.4  \pm 0.6$  &$2789.2  \pm 3.2 $  &
$2791.9   \pm 3.3$  \\
$M(3/2^-)$ & $1519.5   \pm 1.0$    & $2628.1  \pm 0.6$  &$2816.5  \pm 1.2 $  &
$2818.2   \pm 2.1$  \\ \hline
\hline
\end{tabular}
\end{center}
\end{table}

In the heavy quark limit, the ($1/2^-$) and ($3/2^-$) $\Lambda^*$ and $\Xi^*$
excitations listed in Table \ref{tab:orbital}
can be interpreted as a P-wave isospin-0 spinless diquark coupled to the heavy
quark.  Under this assumption, the difference between the spin averaged mass
of the $\Lambda^*$ baryons and the ground state $\Lambda$ is only the orbital
excitation energy of the diquark.
\vbox{
\begin{eqnarray}
\label{eqn:lex}
\Delta E_L(\Lambda) \equiv
\frac{2\Lambda^*_{[3/2]} +\Lambda^*_{[1/2]}}{3}  -\Lambda    &=&366.15\pm1.49~\textnormal{MeV} \nonumber \\
\Delta E_L(\Lambda_c) \equiv
\frac{2\Lambda^*_{c[3/2]}+\Lambda^*_{c[1/2]}}{3} -\Lambda_c  &=&330.74\pm0.47~\textnormal{MeV}  \\
\Delta E_L(\Xi_c) \equiv
\frac{2\Xi^{*}_{c[3/2]}  +\Xi^{*}_{c[1/2]}}{3}  -\Xi_c &=&339.11\pm1.11~\textnormal{MeV} \nonumber
\end{eqnarray}
}

The spin-orbit splitting seems to behave like $1/m_Q$:
\begin{eqnarray}
\label{eqn:fs}
\Lambda^*_{[3/2]}  -\Lambda^*_{[1/2]} &=&113.0\pm4.1 ~\textnormal{MeV} \nonumber \\
\Lambda^*_{c[3/2]} -\Lambda^*_{c[1/2]}&=&\phantom{1}32.7 \pm0.8~\textnormal{MeV}  \\
\Xi^*    _{c[3/2]} -\Xi^*    _{c[1/2]}&=&\phantom{1}26.9\pm2.6~\textnormal{MeV} \nonumber
\end{eqnarray}
where the $\Xi_c$ entries are isospin averages.

The orbital excitation energies in Eq.\ (\ref{eqn:lex}) may be extrapolated to
the case of excited $\Lambda_b$ baryons in the following manner.  Energy
spacings in a power-law potential $V(r) \sim r^\nu$ behave with reduced mass
$\mu$ as $\Delta E \sim \mu^p$, where $p = - \nu/(2+\nu)$ \cite{Quigg:1977dd}.
For light quarks in the confinement regime, one expects $\nu = 1$ and $p =
-1/3$, while for the $c \bar c$ and $b \bar b$ quarkonium states, with nearly
equal level spacings, an effective power is $\nu \simeq 0$ and $p \simeq 0$.
One should thus expect orbital excitations to scale with some power $-1/3
\le p \le 0$.  One can narrow this range by comparing the $\Lambda$
and $\Lambda_c$ excitation energies and estimating $p$ with the help of
reduced masses $\mu$ for the $\Lambda$ and $\Lambda_c$.

\bea
\frac{\mu(\Lambda_c)}{\mu(\Lambda)} =
\frac{M[ud]~m_c}{M[ud] + m_c}\frac{M[ud] + m_s}{M[ud]~m_s}=
\frac{M(\Lambda)}{M(\Lambda_c)}\frac{m_c}{m_s}=1.55
\eea
Now we use the ratio $\Delta E_L(\Lambda_c)/\Delta E_L(\Lambda) = 0.903 \pm
0.004$ to extract an effective power $p = -0.23 \pm 0.01$ which will be used
to extrapolate to the $\Lambda_b$ system:
\bea
\Delta E_L(\Lambda_b) &=&
\Delta E_L(\Lambda_c) \left[ \frac{\mu(\Lambda_b)}{\mu(\Lambda_c)} \right]^p
=\Delta E_L(\Lambda_c)
\left[ \frac{M(\Lambda_c)}{M(\Lambda_b)}\frac{m_b}{m_c} \right]^p
\strut\nonumber\\
\strut\nonumber\\
&=&\Delta E_L(\Lambda_c) 
\left[ \frac{M(\Lambda_c)[M(\Lambda_b)-M(\Lambda)+m_s]}
{M(\Lambda_b)[M(\Lambda_c)-M(\Lambda)+m_s]} \right]^p
\\
\strut\nonumber\\
&=&
\Delta E_L(\Lambda_c)
\left[ \frac{\displaystyle 1-{M(\Lambda)- m_s\over M(\Lambda_b)}}
{\displaystyle 1-{M(\Lambda)- m_s\over M(\Lambda_c)}} \right]^p
 = 317 \pm 1~{\rm MeV}
\nonumber
\label{delta-E-L-Lambda-c}
\eea
where the last form of the expression shows the explicit dependence of the
result on $m_s$.
Using the value $M(\Lambda_b) = (5619.7 \pm 1.2 \pm 1.2)$ MeV observed by
the CDF Collaboration \cite{Acosta:2005mq}, and rescaling the fine-structure
splittings of Eq.\ (\ref{eqn:fs}) by $1/m_Q$ with $m_b/m_c = 2.95 \pm 0.06$,
we find

\begin{equation}
M(\Lambda^*_{b[3/2]}) - M(\Lambda^*_{b[1/2]}) = \frac{m_c}{m_b} (M(\Lambda^*_{c[3/2]})
 - M(\Lambda^*_{c[1/2]})) = (11.1 \pm 0.4)~{\rm MeV}~,
\end{equation}
\begin{equation}
M(\Lambda^*_{b[1/2]}) = (5929 \pm 2)~{\rm MeV}~,~~
M(\Lambda^*_{b[3/2]}) = (5940 \pm 2)~{\rm MeV}~.
\end{equation}

The observed values of the $\Sigma_b$ masses \cite{CDFsigmab},
\begin{eqnarray}
M(\Sigma_b^-) &=& 5815.2{\pm}1.0 ({\rm stat.}) \pm 1.7 ({\rm syst.})
~{\rm MeV} \nonumber \\
M(\Sigma_b^+) &=& {5807.8\,}^{+2.0}_{{-}2.2}\,\,({\rm stat.}) 
\strut\pm\strut 1.7 ({\rm syst.})
~{\rm MeV}
\end{eqnarray}
are sufficiently close to the predicted values of $M(\Lambda^*_{b[1/2,3/2]})$
that the decays
\break
 $\Lambda^*_{b[1/2,3/2]} \to \Sigma_b^\pm \pi^\mp$ \ are
forbidden.  The $\Lambda^*_{b[1/2,3/2]}$ should decay directly to $\Lambda_b
\pi^+ \pi^-$.
\downstrut 

A similar calculation may be performed for the orbitally-excited $\Xi_b$
states.  Here, to a good approximation \cite{Karliner:2007xib}, one may
regard the $[sd]$ diquark in $\Xi_b^-$ or the $[su]$ diquark in $\Xi_b^0$ as
having spin zero, so that methods similar to those applied for excited
$\Lambda_b$ states should be satisfactory.  We find
\beq
\Delta E_L(\Xi_b)
= \Delta E_L(\Xi_c) \left[ \frac{\mu(\Xi_b)}{\mu(\Xi_c)} \right]^p
=\Delta E_L(\Xi_c) \left[ \frac{M(\Xi_c)}{M(\Xi_b)}\frac{m_b}{m_c} \right]^p
= (322 \pm 2)~{\rm MeV}~.
\label{Delta-E-L-Xi-b}
\eeq
Now we use the observed $\Xi_b^-$ mass \cite{CDFxib} $M(\Xi_b^-) = (5792.9
\pm 2.5 \pm 1.7)$ MeV
and our estimate of isospin splitting
$M(\Xi_b^-) - M(\Xi_b^0) = 6.4 \pm 1.6$ MeV to predict the isospin-averaged
value $M(\Xi_b) = 5790 \pm 3$ MeV. We then rescale the fine-structure splitting
(\ref{eqn:fs}) and find
\begin{equation}
\Xi^*_{b[3/2]} - \Xi^*_{b[1/2]} = \frac{m_c}{m_b} (\Xi^*_{c[3/2]}
 - \Xi^*_{c[1/2]}) = (9.1 \pm 0.9)~{\rm MeV}~,
\end{equation}
\begin{equation}
M(\Xi^*_{b[1/2]}) = (6106 \pm 4)~{\rm MeV}~,~~
M(\Xi^*_{b[3/2]}) = (6115 \pm 4)~{\rm MeV}~.
\end{equation}
The lower state decays to $\Xi_b \pi$ via an S-wave, while the higher
state decays to $\Xi_b \pi$ via a D-wave, and hence should be narrower.
Decays to $\Xi'_b \pi$ and $\Xi^*_b \pi$ also appear to be just barely
allowed, given the values of $M(\Xi'_b,\Xi^*_b)$ predicted in Ref.\
\cite{Karliner:2007xib}.

\section{Conclusions}

We have predicted the masses of several baryons containing $b$ quarks, using
descriptions of the color hyperfine interaction which have proved successful
for earlier predictions.  Correcting for wave function effects, we find
$M(\Omega_b) = 6052.1 \pm 5.6$ MeV and $M(\Omega^*_b) = 6082.8 \pm 5.6$ MeV.  
These values are below others which have appeared in
the literature as a result of our use of hadrons containing strange quarks
to evaluate the effective $b-c$ mass difference,
the inclusion of electromagnetic contributions,
and because of a different hyperfine splitting.

We have evaluated the isospin splitting of the $\Xi_b$ states and find
$\Delta I (\Xi_b) \equiv M(\Xi_b^-) - M(\Xi_b^0) = 6.24 \pm 0.21$ MeV on the
basis of an extrapolation from the $\Xi$ and $\Xi^*$ states.  This value is
consistent with one which includes information from the $\Xi_c$ states,
$\Delta I (\Xi_b) = 6.4 \pm 1.6$ MeV.

We have also evaluated the orbital excitation energy for $\Lambda_b$ and
$\Xi_b$ states in which the light diquark ($ud$ or $us$) remains in a
state of $L=S=0$.  Precise predictions have been given for the masses
of the states $\Lambda^*_{b[1/2,3/2]}$ and $\Xi^*_{b[1/2,3/2]}$.
\downstrut

Our predictions are summarized in Table \ref{tab:summary}.  We look forward to
further experimental progress in the tests of these predictions.


\begin{table}[h]
\caption{Summary of predictions for $b$ baryons
}
\label{tab:summary}
\begin{center}
\begin{tabular}{ll} \hline \hline
         & Mass in MeV      \\ \hline
$M(\Omega_b)$   & $6052.1  \pm 5.6$ \\
$M(\Omega_b^*)$ & $6082.8  \pm 5.6$ \\
$M(\Xi^0_{b})$                      & $5786.7  \pm 3.0$ \\
$M(\Lambda^*_{b[1/2]})$               & $5929\phantom{.0} \pm 2$ \\
$M(\Lambda^*_{b[3/2]})$               & $5940\phantom{.0} \pm 2$ \\
$M(\Xi^*_{b[1/2]})$                   & $6106\phantom{.0} \pm 4$ \\
$M(\Xi^*_{b[3/2]})$                   & $6115\phantom{.0} \pm 4$ \\ \hline \hline
\end{tabular}
\end{center}
\end{table}

\section*{Appendix: Values of quark masses}

In choosing values for the quark masses used in this paper,
we note that values of quark mass differences can be taken from the difference in
masses of baryons containing spin-zero $ud$ diquarks
\begin{equation}
m_i - m_j = M(\Lambda_i) - M(\Lambda_j)
\end{equation}
where $i$ and $j$ can be $b, c$ or $s$. This gives
\vskip-2.5em
\bea
&& 
\phantom{aaaaaaaaaaaaaaaaaaaaaaaaaaaaaaaaaaaaaaaaaaaaaaaaaaaaaaaaaaaaaaa}
\nonumber\\ 
m_c &=& M(\Lambda_c) - M(\Lambda)   + m_s=(2286.5 - 1115.68 +538)~{\rm MeV}=
1709~{\rm MeV} 
\nonumber \\
 m_b &=& M(\Lambda_b) - M(\Lambda_c) + m_c \pm 10
~{\rm MeV} =M(\Lambda_b) -  M(\Lambda)   + m_s \pm 10~{\rm MeV}
 \\
&=& 5619.7 - 1115.68 +538\pm 10~{\rm MeV} =
5042\pm 10 ~{\rm MeV}
\nonumber \label{mb-and-mc}
\eea
where $m_s=538~{\rm MeV}$ has been taken from the fit of
Ref.~\cite{Gasiorowicz:1981jz} to light-quark baryon spectra.

We have noted~\cite{Karliner:2003sy} that an uncertainty of 10 MeV arises from
the difference between the values of $m_b - m_c$ obtained from hadrons having
strange and nonstrange spectators. We have chosen the value obtained from
strange spectators following its use in previous successful
predictions~\cite{Karliner:2007xib}.

Although this difference is crucial in predictions like  
Eq.~(\ref{Omegab-spin-ave}) which depend on mass differences, its effect on mass
ratios is negligible. We therefore use the values obtained from baryons with
nonstrange spectators to obtain a value for the mass ratio $m_b/m_c$.

\begin{equation}
\frac{m_b}{m_c}=  \frac
{[M(\Lambda_b)-M(\Lambda)+m_s+\delta m]}{[M(\Lambda_c)-M(\Lambda)+m_s+\delta m]}
=\frac{5042+\delta m}{1709+\delta m}
\approx 2.95 - \frac{\delta m}{876~{\rm MeV}}~,
\end{equation}
where we have introduced the quantity $\delta m$ to take care of any errors in
the assumption that $m_s=538~{\rm MeV}$ and neglected the $10 ~{\rm MeV}$
uncertainty in $m_b$. Taking $\delta m = \pm 50~{\rm MeV}$
in the calculation of $m_b/m_c$ gives the value $m_b/m_c = 2.95 \pm 0.06 $ used in the
$\Omega_b$ mass prediction.

\section*{Acknowledgments}
This research was supported in part by a grant from Israel Science Foundation
administered by Israel Academy of Science and Humanities. The research of
H.J.L. was supported in part by the U.S. Department of Energy, Division of
High Energy Physics, Contract DE-AC02-06CH11357.  Part of this work was
performed while J.L.R. was at the Aspen Center for Physics.
The work of J.L.R. was supported by the U.S. Department of Energy, Division of
High Energy Physics, Grant No.\ DE-FG02-90ER40560.

\end{document}